\documentclass{article}
\usepackage[margin=2cm]{geometry}
\usepackage{amsmath,amssymb,amsfonts,amsthm,graphicx,tikz}

\def\ut#1{\underset{\text{$\tilde{\phantom{\cdot}}$}}{#1}}
\def\uh#1{\underset{\text{$\hat{\phantom{\cdot}}$}}{#1}}
\def\ud#1{\underset{\text{$\dot{\phantom{\cdot}}$}}{#1}}

\def\uth#1{\underset{\hat{\text{{\tiny $\sim$}}}}{#1}}
\def\uhd#1{\underset{\text{$\hat{\cdot}$}}{#1}}
\def\utd#1{\underset{\text{$\tilde{\cdot}$}}{#1}}
\def\uthd#1{\underset{\text{$\hat{\tilde{\cdot}}$}}{#1}}

\def\uthh#1{\underset{\text{$\tilde{\hat{\hat}}$}}{#1}}
\def\utth#1{\underset{\text{$\tilde{\tilde{\hat}}$}}{#1}}
\def\uttd#1{\underset{\text{$\tilde{\tilde{\dot}}$}}{#1}}
\def\utdd#1{\underset{\text{$\tilde{\dot{\dot}}$}}{#1}}
\def\uhdd#1{\underset{\text{$\hat{\dot{\dot}}$}}{#1}}
\def\uhhd#1{\underset{\text{$\hat{\hat{\dot}}$}}{#1}}

\newcounter{NN}
\newtheorem{proposition}[NN]{Proposition}

\newtheorem{corollary}[NN]{Corollary}

\begin{document}

\title{On the relation between the dual AKP equation and an equation by King and Schief, and its N-soliton solution}

\author{
Peter H.~van der Kamp$^1$, Da-jun Zhang$^2$, G.R.W. Quispel$^1$\\[2mm]
$^1$Department of Mathematical and Physical Sciences, La Trobe University, Victoria 3086, Australia\\
$^2$Department of Mathematics, Shanghai University, Shanghai 200444, China\\[5mm]
Email: P.vanderKamp@LaTrobe.edu.au
}

%\date{}

\maketitle

\begin{abstract}
The Dual of the lattice AKP (DAKP) equation [P.H. van der Kamp et al., J. Phys. A 51, 365202 (2018)] is equivalent to a 14-point equation related to the lattice BKP equation, found by King and Schief (KS), and hence it is integrable. We show that the KS equation is dual to Hirota's version of AKP (DAGTE), and we settle the conjectured $N$-soliton solution of DAKP, for all values of the parameters.
\end{abstract}

\section{Introduction}
Each of the three continuous 3D Kadomtsev-Petviashvili equations, called AKP, BKP and CKP, has a discrete counterpart. The first lattice KP equation
\begin{equation} \label{DAGTE}
A\tilde{\tau}\ut{\tau}+
B\hat{\tau}\uh{\tau}+
C\dot{\tau}\ud{\tau}=0,
\end{equation}
was proposed by Hirota \cite{Hir} as a Discrete Analogue of a Generalised Toda Equation (DAGTE). It's cubical form
\begin{equation} \label{AKP}
A\tilde{\tau}\hat{\dot{\tau}}+
B\hat{\tau}\tilde{\dot{\tau}}+
C\dot{\tau}\hat{\tilde{\tau}}=0,
\end{equation}%\cite[Eq. (8.29)]{HJN}
is known as the Hirota-Miwa equation, or the lattice AKP equation. The two equations are related by a linear transformation of the independent variable, i.e. if $\tau=\tau(k,l,m)$ is a solution of equation \eqref{DAGTE} then $\tau(k',l',m')$, where
\begin{equation} \label{LT}
(k,l,m)'=(l+m,k+m,k+l),
\end{equation}
is a solution of equation \eqref{AKP}, cf. \cite[Figure 8.3]{HJN} for a diagrammatic representation. Miwa \cite{Miw} also found the lattice BKP equation (a.k.a. the Miwa equation)
\begin{equation} \label{BKP}
A\tilde{\tau}\hat{\dot{\tau}}+
B\hat{\tau}\tilde{\dot{\tau}}+
C\dot{\tau}\hat{\tilde{\tau}}+
D\tau\hat{\tilde{\dot{\tau}}}=0.
\end{equation}
The lattice CKP equation
was derived by Kashaev \cite{Kas} as a 3D lattice model associated with the local Yang-Baxter relation, and was independently found by Schief \cite{Sch} as the superposition principle for the continuous CKP equation.

Using conservation laws for the lattice AKP equation given in \cite{MQ}{}, the equation\footnote{For functions $\tau(k,l,m)$, which depend on three independent discrete variables, we denote shifts in $k$ using tildes, shifts in $l$ by hats, and shifts in $m$ by dots, e.g. $\underset{\dot{ }}{\hat{\tilde{\tau}}}=\tau(k+1,l+1,m-1)$.
}
\begin{equation} \label{DAKP}
%\begin{split}
a_1\left(\dfrac{\ut{\hat{\dot{\tau}}}}{\hat{\dot{\tau}}\hat{\tau}\dot{\tau}}-
\dfrac{\tilde{\tilde{\tau}}}{\tilde{\tau}\tilde{\hat{\tau}}\tilde{\dot{\tau}}}\right)
+
a_2
\left(\dfrac{\uh{\tilde{\dot{\tau}}}}{\tilde{\dot{\tau}}\tilde{\tau}\dot{\tau}}-
\dfrac{\hat{\hat{\tau}}}{\hat{\tau}\hat{\dot{\tau}}\tilde{\hat{\tau}}}\right)
+
a_3
\left(
\dfrac{\ud{\tilde{\hat{\tau}}}}{\tilde{\hat{\tau}}\tilde{\tau}\hat{\tau}}-
\dfrac{\dot{\dot{\tau}}}{\dot{\tau}\tilde{\dot{\tau}}\hat{\dot{\tau}}}\right)
+
a_4
\left(\dfrac{\tau}{\tilde{\tau}\hat{\tau}\dot{\tau}}-
\dfrac{\tilde{\hat{\dot{\tau}}}}{\tilde{\hat{\tau}}\hat{\dot{\tau}}\tilde{\dot{\tau}}}
\right)=0
%\end{split}
\end{equation}
was derived as a Dual to the AKP equation (DAKP) in \cite{KQZ}. Generally speaking, dual equations to integrable equations do not need to be integrable themselves; the only thing that is guaranteed is the existence
of integrals (for O$\Delta$Es) \cite{QCR}, or conservation laws (for P$\Delta$Es) \cite{KQZ}. In this particular case, DAKP was suspected to be integrable for the following reasons: reductions of DAKP include Rutishauser’s quotient-difference (QD)
algorithm, the higher analogue of the discrete time Toda (HADT) equation and
its corresponding quotient–quotient-difference (QQD) system, the discrete
hungry Lotka–Volterra system, discrete hungry QD, as well as the hungry
forms of HADT and QQD, it was observed that reductions have the Laurent
property and vanishing algebraic entropy, and it was conjectured \cite[Conjecture 1]{KQZ} that the DAKP equation (\ref{DAKP}) admits an $N$-soliton solution of the form
\begin{equation} \label{NSol}
\tau_{k,l,m}=\sum_{w \in P(N)} \Big(\prod_{v\in P_2(w)}R_v\Big) c_w x_w^k y_w^l z_w^m,
\end{equation}
in which $P(N)$ denotes the power set of the string $12\ldots N$, $P_2(S)$ is the subset of the power set of a string $S$ containing all 2-letter substrings, and $x_{ij}=x_ix_j$, $c_{ij}=c_ic_j$. The parameters $x_i,y_i,z_i$ should satisfy the dispersion relations
\begin{equation} \label{dispr}
\begin{split}
Q(x_i,y_i,z_i):=&y_iz_i ( x_i-1 )  ( x_i-y_i )  ( x_i-z_i ) a_{{1}}
+x_iz_i ( y_i-1 )  ( y_i-x_i )  ( y_i-z_i ) a_{{2}}\\
&+x_iy_i ( z_i-1 )
 ( z_i-x_i )  ( z_i-y_i ) a_{{3}}
+x_iy_iz_i ( x_i-1 )  ( y_i-1
 )  ( z_i-1 ) a_{{4}}=0,
\end{split}
\end{equation}
the $c_i$ are constants and the interaction term is given by
\begin{equation} \label{RDAKP}
R_{ij}:=
\frac{a_1S^{ij}_1+a_2S^{ij}_2+a_3S^{ij}_3+a_4S^{ij}_4}{
Q(x_{ij},y_{ij},z_{ij})},
\end{equation}
where
\begin{equation} \label{Sij123}
\begin{split}
S^{ij}_1:=&\big(
( x_i-x_j )  ( x_jy_i-y_jx_i )  ( x_{ij}-z_{ij} )
+ ( x_i-x_j )  ( x_jz_i-z_jx_i ) ( x_{ij}-y_{ij} )\\
&+ ( x_jz_i-z_jx_i )  ( x_jy_i-y_jx_i )  ( 1-x_{ij} )
\big)y_{ij}z_{ij},
\end{split}
\end{equation}
the $S^{ij}_k$ for $k=2$ (respectively $3$) are obtained from $S^{ij}_1$ by interchanging the symbols $x,y$ (respectively $x,z$), and
\begin{equation} \label{Sij4}
\begin{split}
S^{ij}_4:=&\big(
(1 - x_{ij} ) ( y_i-y_j ) ( z_i-z_j )
+ ( x_i-x_j ) (1 - y_{ij} ) ( z_i-z_j ) \\
&+ ( x_i-x_j ) ( y_i-y_j ) (1 - z_{ij} )
\big) x_{ij}y_{ij}z_{ij}.
\end{split}
\end{equation}

Schief \cite{SPC} pointed out that (\ref{DAKP}), with $a_1=a_2=a_3=-a_4=1$, is related to a 14-point equation in \cite{KS}. King and Schief remarkably observed this equation to be a consequence of the BKP equation, with coefficients $\pm1$. A consequence of BKP \eqref{BKP} with general coefficients is given by
\begin{equation} \label{KS}
A^2 \left( \frac{\uhd{\tilde{\tau}}}{\tilde{\tau}\uh{\tau}\ud{\tau}}
-\frac{\ut{\hat{\dot{\tau}}}}{\ut{\tau}\hat{\tau}\dot{\tau}}\right)+
B^2 \left(\frac{\utd{\hat{\tau}}}{\ut{\tau}\hat{\tau}\ud{\tau}}
-\frac{\uh{\tilde{\dot{\tau}}}}{\tilde{\tau}\uh{\tau}\dot{\tau}}\right)+
C^2 \left(\frac{\uth{\dot{\tau}}}{\ut{\tau}\uh{\tau}\dot{\tau}}
-\frac{\ud{\hat{\tilde{\tau}}}}{\tilde{\tau}\hat{\tau}\ud{\tau}}\right)+
D^2 \left(\frac{\hat{\tilde{\dot{\tau}}}}{\tilde{\tau}\hat{\tau}\dot{\tau}}
-\frac{\uthd{\tau}}{\ut{\tau}\uh{\tau}\ud{\tau}}
\right) = 0.
\end{equation}
and will be called the KS equation. The observation by King and Schief is now well-understood, as equation \eqref{KS} arises as a reduction of the multi-component extension of BKP \cite[Eq. (2.45)]{ZVZ},
\begin{equation}\label{2111B}
\begin{split}
A\tilde{\tau}\hat{\dot{\sigma}}+
B\hat{\tau}\tilde{\dot{\sigma}}+
C\dot{\tau}\hat{\tilde{\sigma}}+
D\sigma\hat{\tilde{\dot{\tau}}}&=0\\
A\tilde{\sigma}\hat{\dot{\tau}}+
B\hat{\sigma}\tilde{\dot{\tau}}+
C\dot{\sigma}\hat{\tilde{\tau}}+
D\tau\hat{\tilde{\dot{\sigma}}}&=0,
\end{split}
\end{equation}
which is the 2[1,1,1] extension. Elimination of $\sigma$ from \eqref{2111B} yields \eqref{KS}, for details we refer to \cite[Section 3.2]{ZVZ}. 

The relation between the DAKP equation \eqref{DAKP} and the KS equation \eqref{KS} (identifying $a_1=A^2,a_2=B^2,a_3=C^3,a_4=-D^2$) is given by the same linear transformation of the independent variables that relates the AKP equation \eqref{AKP} to the DAGTE \eqref{DAGTE}. Not only does this establish the integrability of DAKP, it also implies that the KS equation is a dual to the DAGTE. A diagram of the relations between the various equations is given in Fig. \ref{Dia}.
\begin{figure}[h!]
\centering
\begin{tikzpicture}[scale=1]
\tikzstyle{nod}= [draw, rounded corners]		
\node[nod] (a) at (0,0) {DAKP};
\node[nod] (b) at (0,2) {DAGTE};
\node[nod] (c) at (4,0) {AKP};
\node[nod] (d) at (4,2) {KS};
\node[nod] (e) at (8,2) {2[1,1,1] BKP};
\node[nod] (f) at (8,0) {BKP};
\node at (0,1) {transformation \eqref{LT}};
%\node at (3.7,1) {\eqref{LT}};
\node at (2,-.3) {dual};
\node at (2,2.3) {dual};
\node at (5.7,2.3) {reduction};
\node at (5.7,-.3) {$D=0$};
\node at (10,1) {2-component extension};
\draw[<->, line width=1pt] (c) -- (b);
\draw[<->, line width=1pt] (c) -- (a);
\draw[<->, line width=1pt] (b) -- (d);
\draw[<->, line width=1pt] (a) -- (d);
\draw[->, line width=1pt] (e) -- (d);
\draw[->, line width=1pt] (f) -- (e);
\draw[->, line width=1pt] (f) -- (c);
\end{tikzpicture}
\caption{\label{Dia} Relations between 6 equations.}
\end{figure}

In this short note, we provide explicitly the matrix conservation law that establishes the duality between the DAGTE and the KS equation and we prove Conjecture 1 in \cite{KQZ}, i.e. that the conjectured equation for the soliton solution of DAKP is correct. The proof is not entirely straightforward; the fact that the coefficients $a_1,a_2,a_3,a_4$ are general means in particular that they may vanish. However, in the Miwa-parametrisation of the soliton solution for the BKP equation one can not set a single coefficient to zero. If one of the coefficients vanishes, the DAKP soliton solution actually relates to an AKP soliton solution, which is a solution of a related 12-point equation \cite[Eq. (3.18)]{ZVZ}.

\section{Transformation of the independent variables}
\begin{figure}[h]
\hspace{2cm} \includegraphics[width=5cm]{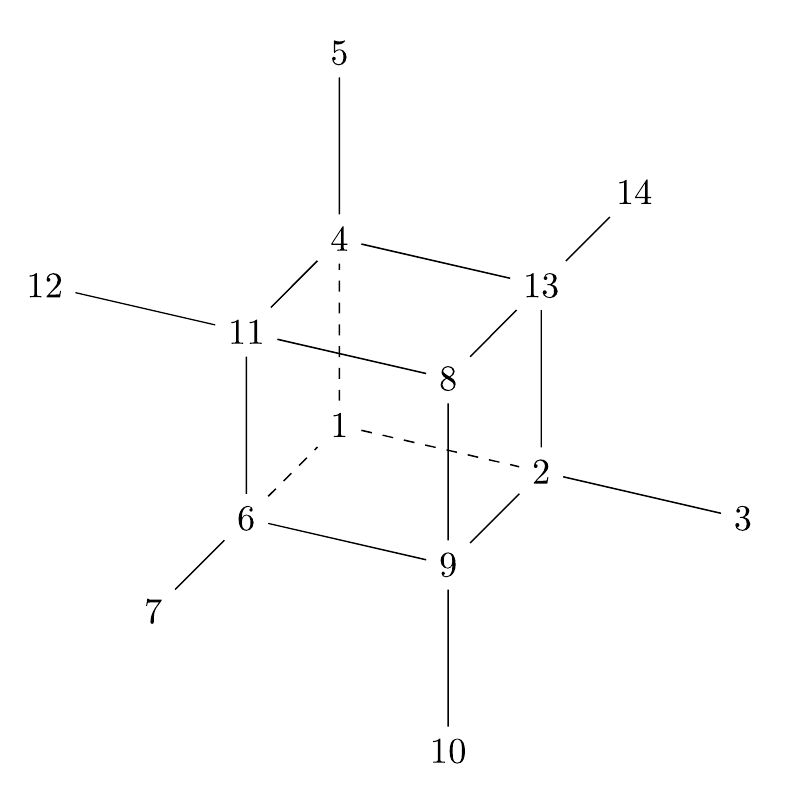} \hspace{2cm} \includegraphics[width=5cm]{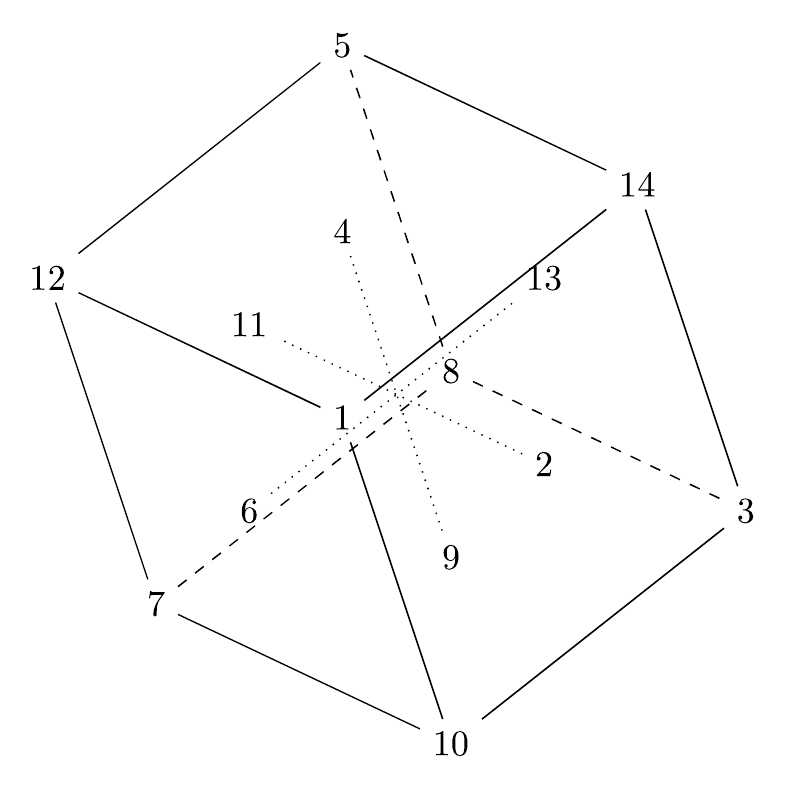}
\caption{\label{STC} The lattice points, represented by numbers 1 to 14, appear in the same spots in both diagrams.
On the left the 14-point stencil of DAKP, where $\tau=1,\tilde{\tau}=2,\hat{\tau}=4,\dot{\tau}=6$, and on the right the 14-point stencil of KS, where $\tau=1,\tilde{\tilde{\tau}}=12,\hat{\hat{\tau}}=10,
\dot{\dot{\tau}}=14$. On the left $2=\tilde{\tau}$ is a single shift, whereas at on the right the point $2=\hat{\dot{\tau}}$ appears in the center of the face $1-10-3-14$. \hfill $\square$}
\end{figure}

The relation between DAKP and KS (upshifted in each direction) is given as follows \cite{KS}.
\begin{proposition} \label{rel2}
For all $A,B,C,D$, if $\tau=\tau(k,l,m)$ is a solution of Eq. (\ref{KS}) then $\tau(l+m,k+m,k+l)$ is a solution of Eq. (\ref{DAKP}) with $a_1=A^2,a_2=B^2,a_3=C^3,a_4=-D^2$.
\end{proposition}
\noindent
{\bf Proof.} Changing independent variables, the (stencils of the) equations are transformed into each other, see Fig. \ref{STC}.

\begin{figure}[h]
\begin{center}
\parbox{8cm}{\includegraphics[width=5.7cm]{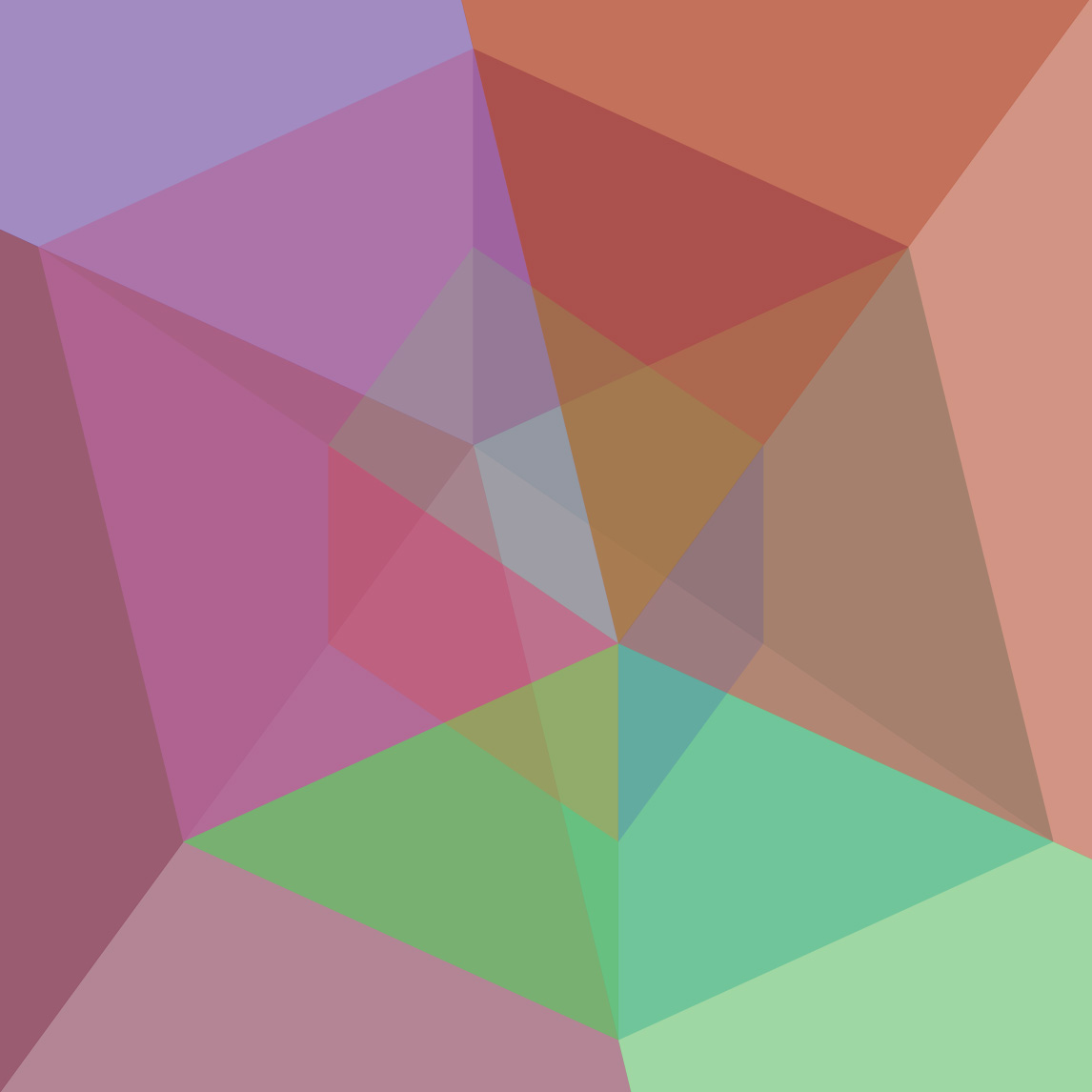}} 
\qquad \parbox{6cm}{\caption{\label{TRANS} {\em Transformational, by Peter H van der Kamp.} A lattice on a lattice, a cube inside a cube, in our minds we change what is back to front.}}
\end{center}
\end{figure}

A more artistic representation of the transformation, combining both diagrams of Fig. \ref{STC} is given in Figure \ref{TRANS}.

\section{Duality between DAGTE and KS, a matrix conservation law}
\begin{proposition} \label{duality}
The KS equation \eqref{KS} is a dual to the DAGTE \eqref{DAGTE}.
\end{proposition}
\noindent
{\bf Proof.} Applying \eqref{LT} to \cite[Eq. (5)]{KQZ} gives us the following equation,
\begin{equation} \label{PCONS}
\dot{\hat{{\cal P}}}-{\cal P}+\dot{\tilde{{\cal Q}}}-{\cal Q}+\hat{\tilde{{\cal R}}}-{\cal R}={\cal V}^T{\cal W},
\end{equation}
where
\[
{\cal P}=\left[\begin{array}{cccc}
-\frac{\ut{\tau} \tilde{\uhd{\tau}}}{\ud{\tau} \uh{\tau}} & 0 & 0 & -\frac{\tilde{\tau} \uthd{\tau}}{\ud{\tau} \uh{\tau}} 
\\
 -\frac{\ut{\tau} \uhhd{\tau}}{\uh{\tau} \uthd{\tau}} & 0 & \frac{\ud{\tau} \uthh{\tau}}{\uh{\tau} \uthd{\tau}} & 0 
\\
 -\frac{\uhdd{\tau} \ut{\tau}}{\uthd{\tau} \ud{\tau}} & \frac{\utdd{\tau} \uh{\tau}}{\uthd{\tau} \ud{\tau}} & 0 & 0 
\end{array}\right],\quad
{\cal Q}=\left[\begin{array}{cccc}
0 & -\frac{\uttd{\tau} \uh{\tau}}{\uthd{\tau} \ut{\tau}} & \frac{\ud{\tau} \utth{\tau}}{\uthd{\tau} \ut{\tau}} & 0 
\\
 0 & -\frac{\uh{\tau} \hat{\utd{\tau}}}{\ud{\tau} \ut{\tau}} & 0 & -\frac{\hat{\tau} \uthd{\tau}}{\ud{\tau} \ut{\tau}} 
\\
 \frac{\uhdd{\tau} \ut{\tau}}{\uthd{\tau} \ud{\tau}} & -\frac{\utdd{\tau} \uh{\tau}}{\uthd{\tau} \ud{\tau}} & 0 & 0 
\end{array}\right],\quad
{\cal R}=\left[\begin{array}{cccc}
0 & \frac{\uttd{\tau} \uh{\tau}}{\uthd{\tau} \ut{\tau}}
& -\frac{\ud{\tau} \utth{\tau}}{\uthd{\tau} \ut{\tau}} & 0 
\\
 \frac{\ut{\tau} \uhhd{\tau}}{\uh{\tau} \uthd{\tau}} & 0 & -\frac{\ud{\tau} \uthh{\tau}}{\uh{\tau} \uthd{\tau}} & 0 
\\
 0 & 0 & -\frac{\ud{\tau} \uth{\dot{\tau}}}{\ut{\tau} \uh{\tau}} & -\frac{\dot{\tau} \uthd{\tau}}{\ut{\tau} \uh{\tau}} 
\end{array}\right]
\]
and
\[
{\cal V}=\left(\tilde{\tau} \ut{\tau},
 \uh{\tau} \hat{\tau},
 \ud{\tau} \dot{\tau}\right),\quad
{\cal W}=
\left(\frac{\tilde{\uhd{\tau}}}{\tilde{\tau} \ud{\tau} \uh{\tau}}
-\frac{\dot{\hat{\ut{\tau}}}}{\ut{\tau} \hat{\tau} \dot{\tau}},
\frac{\hat{\utd{\tau}}}{\hat{\tau} \ud{\tau} \ut{\tau}}
-\frac{\dot{\tilde{\uh{\tau}}}}{\uh{\tau} \tilde{\tau} \dot{\tau}},
\frac{\uth{\dot{\tau}}}{\dot{\tau} \ut{\tau} \uh{\tau}}
-\frac{\hat{\tilde{\ud{\tau}}}}{\ud{\tau} \hat{\tau} \tilde{\tau}},
\frac{\uthd{\tau}}{\ut{\tau} \ud{\tau} \uh{\tau}}
-\frac{\dot{\hat{\tilde{\tau}}}}{\tilde{\tau} \hat{\tau} \dot{\tau}} 
\right).
\]
Equation \eqref{PCONS} can be written as a conservation law as follows:
\begin{equation} \label{CONS}
\widetilde{\hat{\cal R}+{\cal Q}} - (\hat{\cal R}+{\cal Q})+
\widehat{\dot{\cal P}+{\cal R}} - (\dot{\cal P}+{\cal R})+
\overset{\text{{\tiny \textbullet}}}{(\tilde{\cal Q}+{\cal P})} - (\tilde{\cal Q}+{\cal P})
={\cal V}^T{\cal W}.
\end{equation}
Denoting vectors of coefficients ${\bf{v}}=\left( A , B , C \right)$ and ${\bf{w}}=\left(A^2, B^2, C^2, D^2 \right)$, we have that ${\bf{v}}{\cal V}^T=0$ represents the DAGTE (\ref{DAGTE}) and the equation ${\cal W}{\bf{w}}^T=0$ is the KS equation (\ref{KS}). Pre-multiplying (\ref{CONS}) with ${\bf{v}}$  gives four conservation laws for DAGTE, and post-multiplying (\ref{CONS}) with ${\bf{w}}^T$ yields three conservation laws for the KS equation. \hfill $\square$

\section{The soliton solution of DAKP}
It is well known that both the lattice AKP equation \eqref{AKP},
with $A+B+C=0$, and the lattice BKP equation  \eqref{BKP}, with $A+B+C+D=0$, have an $N$-soliton solution of the form (\ref{NSol}). These $N$-soliton solutions have been parametrised by Miwa \cite{Miw} as follows.
\begin{itemize}
\item
For the AKP equation one sets
\begin{equation} \label{ABC}
A= a \left( b-c \right),\
B= b\left( c-a \right),\
C= c\left( a-b \right)
\end{equation}
and
\[
x_{{i}}={\frac {1-aq_{{i}}}{1-ap_{{i}}}},\
y_{{i}}={\frac {1-bq_{{i}}}{1-bp_{{i}}}},\
z_{{i}}={\frac {1-cq_{{i}}}{1-cp_{{i}}}}.
\]
The AKP interaction term is given in terms of the $p_i,q_i$ by
\begin{equation} \label{RAKP}
R_{ij} = -{\frac { \left( q_{{i}}-q_{{j}} \right)  \left( p_{{i}}-p_{{j}}
 \right) }{ \left( p_{{j}}-q_{{i}} \right)  \left( p_{{i}}-q_{{j}}
 \right) }}.
\end{equation}
\item
For the BKP equation the parameters are
\begin{equation} \label{ABCD}
\begin{split}
A&= \left( b+a \right)  \left( c+a \right)  \left( b-c \right),\
B= \left( c+b \right)  \left( b+a \right)  \left( c-a \right),\\
C&= \left( c+a \right)  \left( c+b \right)  \left( a-b \right),\
D= \left( a-b \right)  \left( b-c \right)  \left( c-a \right),
\end{split}
\end{equation}
and
\begin{equation}
x_i={\frac { \left( 1-ap_i \right)  \left( 1-aq_i \right)}{ \left( ap_i+1 \right)  \left( aq_i+1 \right) }},\
y_i={\frac { \left( 1-bp_i \right)  \left( 1-bq_i \right) }{ \left( bp_i+1 \right)
 \left( bq_i+1 \right) }},\
z_i={\frac { \left( 1-cp_i \right)  \left( 1-cq_i \right) }{ \left( cp_i+1 \right)  \left( cq_i+1 \right) }}.
\end{equation}
The BKP interaction term is given in terms of the $p_i,q_i$ by
\begin{equation} \label{RBKP}
R_{ij}=
{\frac { \left( q_{{i}}-q_{{j}} \right)  \left( q_{{i}}-p_{{j}}
 \right)  \left( p_{{i}}-q_{{j}} \right)  \left( p_{{i}}-p_{{j}}
 \right) }{ \left( q_{{i}}+q_{{j}} \right)  \left( q_{{i}}+p_{{j}}
 \right)  \left( p_{{i}}+q_{{j}} \right)  \left( p_{{i}}+p_{{j}}
 \right) }}.
\end{equation}
\end{itemize}

We aim to show that the conjectured equation for the soliton solution of DAKP is correct, for all values of the coefficients $a_1,a_2,a_3,a_4$, including cases where one or more of them vanish. This implies we also need to work with the 12-point equation
\begin{equation} \label{12p}
A^2 \left( \frac{\uhd{\tilde{\tau}}}{\tilde{\tau}\uh{\tau}\ud{\tau}}
-\frac{\ut{\hat{\dot{\tau}}}}{\ut{\tau}\hat{\tau}\dot{\tau}}\right)+
B^2 \left(\frac{\utd{\hat{\tau}}}{\ut{\tau}\hat{\tau}\ud{\tau}}
-\frac{\uh{\tilde{\dot{\tau}}}}{\tilde{\tau}\uh{\tau}\dot{\tau}}\right)+
C^2 \left(\frac{\uth{\dot{\tau}}}{\ut{\tau}\uh{\tau}\dot{\tau}}
-\frac{\ud{\hat{\tilde{\tau}}}}{\tilde{\tau}\hat{\tau}\ud{\tau}}\right)
= 0,
\end{equation}
which is a special case of the KS equation with $D=0$, and a reduction of the 2[1,1,1] extension of AKP \cite[Eq. (2.44)]{ZVZ}\footnote{By a general result, multi-component generalisations are multi-dimensionally consistent and they provide B\"acklund transformations/a Lax pair for their reductions \cite{ZVZ}.},
\begin{equation}\label{2111A}
\begin{split}
A\tilde{\tau}\hat{\dot{\sigma}}+
B\hat{\tau}\tilde{\dot{\sigma}}+
C\dot{\tau}\hat{\tilde{\sigma}}&=0\\
A\tilde{\sigma}\hat{\dot{\tau}}+
B\hat{\sigma}\tilde{\dot{\tau}}+
C\dot{\sigma}\hat{\tilde{\tau}}&=0.
\end{split}
\end{equation}

To establish a connection between the $N$-soliton solution for DAKP and the above $N$-soliton solutions of AKP and BKP, we make use the following version of \cite[Theorem 3.2]{KS}.
\begin{proposition} \label{rel1}
For all $A,B,C,D$, if $\tau$ satisfies the BKP equation, then $\tau$ satisfies equation \eqref{KS}.
\end{proposition}
\noindent
{\bf Proof.} Similar to the proof given by King and Schief, shifting the BKP equation backwards in all directions, dividing by $\tau$ and the taus containing one shift, and taking a linear combination, this time allowing the coefficients to be $\pm A,\pm B,\pm C$, and $\pm D$, yields the result. Explicitly, defining $E$ to be the left hand side of (\ref{BKP}), we have that
\begin{equation*}
\frac{A}{\tilde{\tau}\uh{\tau}\ud{\tau}}\uhd{E}
+\frac{B}{\ut{\tau}\hat{\tau}\ud{\tau}}\utd{E}
+\frac{C}{\ut{\tau}\uh{\tau}\dot{\tau}}\uth{E}
+\frac{D}{\tilde{\tau}\hat{\tau}\dot{\tau}}{E}
-\frac{A}{\ut{\tau}\hat{\tau}\dot{\tau}}\ut{E}
-\frac{B}{\tilde{\tau}\uh{\tau}\dot{\tau}}\uh{E}
-\frac{C}{\tilde{\tau}\hat{\tau}\ud{\tau}}\ud{E}
-\frac{D}{\ut{\tau}\uh{\tau}\ud{\tau}}\uthd{E}
\end{equation*}
equals the left hand side of (\ref{KS}), multiplied by $\tau$. \hfill $\square$

Note that the parameters are not essential, with $D\neq0$ one can absorb $D^2$ and then perform the transformation, cf. equation (2.5) in  \cite{Nim},
\begin{equation} \label{transf}
\tau(k,l,m)\rightarrow \left(\frac{BC}{A}\right)^{lm}
\left(\frac{AC}{B}\right)^{km}
\left(\frac{AB}{C}\right)^{kl}\tau(k,l,m),
\end{equation}
that is, if the parameters $A,B,C,D$ are not zero they can be set to 1, or any other nonzero value. Taking $D=0$ yields the following corollary.
\begin{corollary} \label{cor}
For all $A,B,C$, if $\tau$ satisfies the AKP equation, then $\tau$ satisfies the 12-point equation \eqref{12p}.
\end{corollary}
Of course, the same transformation of the independent variables provides the relation between the DAKP equation (\ref{DAKP}) with $a_4=0$ and the 12-point equation (\ref{12p}), where $A^2=a_1$, $B^2=a_2$, $C^2=a_3$. We will now show how the $N$-soliton solution of the DAKP equation relates to the $N$-soliton solutions of the AKP and BKP equations.
\begin{proposition} \label{NS}
The $N$-soliton solution of DAKP, given by Eqs. \eqref{NSol}-\eqref{Sij4}, can be formulated using Miwa's parametrisation.
\end{proposition}

\noindent {\bf Proof}
\begin{itemize}
\item For nonzero $a_1,a_2,a_3,a_4$  one can apply a transformation similar to (\ref{transf}), which induces a transformation on the parameters $a_i\rightarrow a_i^\prime$ ($i=1,2,3$) and $a_4\rightarrow a_4^\prime$ can be scaled independently, such that after the transformation we can parametrise
\begin{align*}
a_1^\prime&= \big(\left( b+a \right)  \left( c+a \right)  \left( b-c
 \right) \big)^2,\quad
a_3^\prime= \big(\left( c+a \right)  \left( c+b \right)  \left( a-b
 \right) \big)^2,\\
a_2^\prime&=\big(\left( c+b \right)  \left( b+a \right)  \left( c-a
 \right) \big)^2,\quad
a_4^\prime= -\big(\left( a-b \right)  \left( b-c \right)  \left( c-a
 \right)\big)^2,
\end{align*}
and
\begin{align*}
x_i&={\frac { \left( 1-bp_i \right)  \left( 1-bq_i \right)\left( 1-cp_i \right)  \left( 1-cq_i \right)}{ \left( bp_i+1 \right)  \left( bq_i+1 \right)\left( cp_i+1 \right) \left( cq_i+1 \right) }},
y_i={\frac { \left( 1-ap_i \right)  \left( 1-aq_i \right)\left( 1-cp_i \right)  \left( 1-cq_i \right)}{ \left( ap_i+1 \right)  \left( aq_i+1 \right)\left( cp_i+1 \right)
 \left( cq_i+1 \right) }},\\
z_i&={\frac { \left( 1-ap_i \right)  \left( 1-aq_i \right)\left( 1-bp_i \right)  \left( 1-bq_i \right)}{ \left( ap_i+1 \right)  \left( aq_i+1 \right)\left( bp_i+1 \right)
 \left( bq_i+1 \right) }}.
\end{align*}
The DAKP dispersion relations $Q(x_i,y_i,z_i)=0$ (with $a_i\rightarrow a_i^\prime$) are then satisfied. Furthermore, the DAKP interaction term $R_{ij}$ (\ref{RDAKP}), in terms of the $p_i,q_i$, is precisely (\ref{RBKP}).

\item Suppose that $a_4=0$. After a transformation (if necessary), parametrising
$a_1^\prime= a^2 \left( b-c \right)^2$,
$a_2^\prime= b^2 \left( c-a \right)^2$,
$a_3^\prime= c^2 \left( a-b \right)^2$,
and
\begin{equation*}
x_{{i}}={\frac {1-bq_{{i}}}{1-bp_{{i}}}}{\frac {1-cq_{{i}}}{1-cp_{{i}}}},\quad
y_{{i}}=
{\frac {1-aq_{{i}}}{1-ap_{{i}}}}{\frac {1-cq_{{i}}}{1-cp_{{i}}}},\quad
z_{{i}}={\frac {1-aq_{{i}}}{1-ap_{{i}}}}{\frac {1-bq_{{i}}}{1-bp_{{i}}}},
\end{equation*}
the DAKP dispersion relations $Q(x_i,y_i,z_i)=0$ are satisfied. Moreover, the DAKP interaction term $R_{ij}$ (\ref{RDAKP}) in terms of the $p_i,q_i$ is precisely (\ref{RAKP}).

\item Suppose that $a_1=0$. Changing the direction of the tilde-shift, i.e. $\tilde{\tau}\rightarrow \ut{\tau}$, the BKP equation with $A=0$ becomes an AKP equation with $A\rightarrow D$, $B\rightarrow C$, and $C\rightarrow B$. Then applying a transformation similar to (\ref{transf}) one can parametrise
$a_2^\prime= c^2 \left( a-b \right)^2$,
$a_3^\prime= b^2 \left( c-a \right)^2$,
$a_4^\prime= - a^2 \left( b-c \right)^2$,
and with
\begin{equation*}
x_{{i}}={\frac {1-bp_{{i}}}{1-bq_{{i}}}}{\frac {1-cq_{{i}}}{1-cp_{{i}}}},\quad
y_{{i}}=
{\frac {1-aq_{{i}}}{1-ap_{{i}}}}{\frac {1-cq_{{i}}}{1-cp_{{i}}}},\quad
z_{{i}}={\frac {1-aq_{{i}}}{1-ap_{{i}}}}{\frac {1-bq_{{i}}}{1-bp_{{i}}}},
\end{equation*}
the DAKP dispersion relations $Q(x_i,y_i,z_i)=0$ are then satisfied. Moreover, the DAKP interaction term $R_{ij}$ (\ref{RDAKP}) in terms of the $p_i,q_i$ is precisely (\ref{RAKP}).

\item Similar parametrisations can be given when $a_2=0$, or $a_3=0$. Changing the direction of the hat-shift transforms BKP with $B=0$ into AKP with $A\rightarrow D$, $B\rightarrow C$ and $C\rightarrow B$. Changing the direction of the dot-shift transforms BKP with $C=0$ into AKP with $A\rightarrow B$, $B\rightarrow A$ and $C\rightarrow D$. \hfill $\square$
\end{itemize}

\section{Concluding remarks}

The KS equation and the DAKP equation have been derived in entirely different settings, their equivalence is quite remarkable. Equally remarkable is the fact that KS equation is on the one hand a dual to DAGTE and on the other a reduction of a two-component extension of BKP.

The DAKP equation was derived as a dual to AKP by making use of four rational characteristics presented in \cite{MQ}, namely $\Lambda_1/A,\Lambda_2/B,\Lambda_3/C$ and $\Lambda_7$, with $D=0$, and constant functions $f_i=1$, $i=1,2,3,7$. However, more (parameter dependent) characteristics were given: six for both AKP and BKP and a seventh for AKP. Hence, a more general dual to AKP is, with $D=0$,
\begin{equation} \label{GDAKP}
\sum_{i=1}^7 \Lambda_i=0,
\end{equation}
and a dual to BKP is, with $D\neq0$,  
\begin{equation} \label{GDBKP}
\sum_{i=1}^6 \Lambda_i=0.
\end{equation}
The question whether any multi-parameter {\em integrable} subclasses of these equations exist, other than DAKP, remains open.  

\subsection*{Acknowledgements}
The authors are indebted to Wolfgang Schief for providing the relation between DAKP and the KS equation. Financial support was provided by a La Trobe University China studies seed-funding research grant, by the department of Mathematics and Statistics of La Trobe University, and by the NSF of China [grant no. 12271334]. GRWQ's work was partially supported by a grant from the Simons Foundation.

\end{document}